\documentclass[prl,twocolumn,preprintnumbers,amsmath,amssymb,floatfix]{revtex4}

\newcommand \be  {\begin{equation}}
\newcommand \beq {\begin{equation}}
\newcommand \bea {\begin{eqnarray} }
\newcommand \ee  {\end{equation}}
\newcommand \eeq {\end{equation}}
\newcommand \eea {\end{eqnarray}}
\newcommand{\beqa}{\begin{eqnarray}}
\newcommand{\eeqa}{\end{eqnarray}}

\usepackage{amsmath, amsfonts, amssymb,graphicx,bm}

\begin{document}                

\title{ 
Topological methods for searching barriers and reaction paths
}  
\author{Sorin T\u{a}nase-Nicola and Jorge Kurchan}
\affiliation{ 
 P.M.M.H. Ecole Sup{\'e}rieure de Physique et Chimie Industrielles,
\\
10, rue Vauquelin, 75231 Paris CEDEX 05,  France}
\date{\today}

\begin{abstract}

We present a family of algorithms for the fast determination of reaction paths 
and barriers in phase space and the computation of the corresponding rates.
 The method requires  the reaction times be large compared to the microscopic time, 
irrespective of the origin --- energetic, entropic, cooperative --- of the timescale separation.
 It lends itself to temperature cycling as in simulated annealing and to activation-relaxation
routines.
The dynamics is ultimately based on  supersymmetry methods used years ago to derive Morse 
theory.
Thus, the formalism  automatically incorporates all relevant topological
information.
 
\end{abstract}
\maketitle

A situation often met in chemical reactions, protein folding and nucleation
transitions is that the evolution breaks into fast local relaxations
and rare {\em activation} processes.
 This timescale separation can be a
 consequence of low temperature
(intra-valley equilibration occurs in 
$\tau_{fast} = O(1)$, passage times are exponential  
$\tau_{activ} \sim \exp(\Delta E/T)$), 
or of cooperative dynamics: e.g. in a 
ferromagnet of size $L$ the relaxation time within a state
$\tau_{fast} < L^2 $ (the time for the largest domain to
 collapse), and between phases
 $\tau_{activ} \sim \exp(cL^{d-1})$. It can also be the result of
 of {\em entropic barriers}, 
flat but narrow energy canyons that the system is unlikely to
 transit.

In all these cases,  a knowledge of the actual connectivity of
paths and saddles gives a global picture of a system's behavior \cite{Wales03}.
 Direct information on the barriers can also  be used to model 
the activated processes, without the need to wait for them to 
occur spontaneously,  thus allowing to simulate a reaction
even faster  than nature.
Furthermore, {\em activation/relaxation} procedures \cite{Mousseau98}, alternating
ordinary relaxation with forced activation, offer an efficient way to sample phase-space.
For these investigations, several algorithms
\cite{Mousseau98,Wales03} have been  implemented to locate saddle
points in energy.
 In cases  when the nature of barriers is qualitatively
 affected by entropy, one needs to study directly
the transition probability --- thus taking into account
the multiplicity of paths through the (``free energy'') barrier. A well known method to do this 
is  a MonteCarlo sampling of the  trajectories with   ends fixed in the initial and
the target state, known as
 `Transition Path Sampling' 
\cite{Dellago98,Bolhuis02,E02}. In practice, this means going from an `open-ended' 
to a `two-ended' procedure.

The purpose of this paper is to present a family of methods for the determination
of  reaction distributions and barriers. It is  based on a dynamics that
 converges to barriers and reaction paths, much in the same way that 
diffusion in  a potential converges to energy minima and metastable states.
This allows us to construct  `free energy barrier' algorithms that can however be implemented
 in an open-ended manner. 

The organization of this paper is as follows: in the first section we present 
the dynamics and we justify it in the second section. 

\noindent {\bf \small{  DIFFUSION DYNAMICS AND PATH SAMPLING}}

Consider a Langevin process in an  $N$-dimensional 
energy landscape $E(x_1,...,x_N)$:
\begin{equation}
{\dot x}_i = - E_i + \eta_i(t)
\label{langevin}
\end{equation}
where $\eta(t)$ is a white noise of variance $2T$.
Here and in what follows $E_i$ and $E_{ij}$ denote
derivatives of $E$, $\partial_i\equiv \frac{\partial}{\partial x_i}$ and we adopt 
the summation convention.
An equivalent way of stating (\ref{langevin})
 is to consider the  probability distribution $P(x,t)$, 
evolving according to:
\begin{eqnarray}
\frac{\partial P(x,t)}{\partial t} 
&=& -H_{\mbox{{\tiny FP}}} P(x,t) =-\partial_i J_i(x)\label{evolution} \\
H_{\mbox{{\tiny FP}}} &=& 
-\partial_i \left(T\partial_i+ E_i \right) \label{HFP}
\label{fokker}
\end{eqnarray} 
which defines  the current
$
J_i(x,t) \equiv \left(T\partial_i+ E_i\right) P(x,t)
$.

The problem is how to modify (\ref{langevin}) in order to devise 
 a diffusive dynamics that targets saddle points and reaction paths.
We shall show below that a set of $N$ functions $R_i({\boldsymbol x})$
evolving with
\begin{equation}
\frac{\partial R_i(x,t)}{\partial t} = 
- H_{\mbox{{\tiny FP}}}  R_i(x,t) - E_{ij}({\boldsymbol x}) R_j(x,t)
\label{onef}
\end{equation}
 yields, at times $\tau_{fast}<t \ll \tau_{activ}$  
a vector field $R_i(t) =J_i({\boldsymbol x})$ describing reaction currents
 between metastable states ---
 which one depends on the initial conditions. 
{\em In other words, (\ref{onef}) does for reaction
paths what (\ref{evolution}) does for states.
}

Equation (\ref{onef}) is not immediately suitable for practical computations. 
Just as the natural way to simulate (\ref{evolution}) is to use a diffusion dynamics  
 (\ref{langevin}), 
in order to have a practical approach we have to find the 
 diffusion dynamics
yielding    (\ref{onef}). 

{\bf Vector walkers.} We consider a population of independent walkers ${\boldsymbol{x^a (t)}}$ 
 carrying a vector internal degree of freedom ${\boldsymbol{v^a (t)}}$,
an $N$-dimensional normalized vector.
The dynamics consists of three types of transformations. Putting  
$A({\boldsymbol x^a}) \equiv - v_i^a E_{ij} v^a_j$, 
they are :
\begin{itemize}
\item Ordinary diffusion of the positions  
${\boldsymbol{x^a (t)}}$ as in (\ref{langevin});
\item Length preserving rotation of the vector with a po\-si\-tion-dependent rate: 
$\frac{dv_i^a}{dt}=-E_{ij} v^a_j(t)-A({\boldsymbol x^a}) v^a_i$;
\item Cloning with rate
 $A({\boldsymbol x^a})$ (if $A({\boldsymbol x^a})>0$),
 or death with rate
$-A({\boldsymbol x^a})$ (if $A({\boldsymbol x^a})<0$). Clones
are born in the same place with the same vector
\footnote{Occasionally one must adjust the total number of walkers with a global death/birth rate.}. 
\end{itemize}

\begin{figure}
\centering \includegraphics[width=70mm]{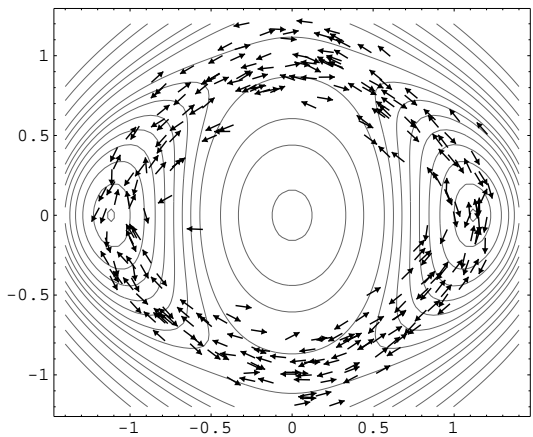}
\caption{Vector walkers  
describing a reaction path. The potential and the temperature
 are the ones of
Ref. \cite{Dellago98}. There are  two minima
  at  $(\pm\sqrt{{5}/{4}},0)$,  
 two saddles at $(0,\pm 1)$ and  a maximum at $(0,0)$.
}
\label{arrowsnapchan}
\end{figure}

The rotation-cloning steps play the role of singling out the least stable direction,  
somewhat like in eigenvector-following and related methods \cite{Wales03, Mousseau98}. 
After a transient $\tau_{fast}< t \ll \tau_{activ}$,
 the system tends to a regime in which
 particles organize along  trails: the reaction paths (Fig. \ref{arrowsnapchan}).
The distribution is almost stationary, 
there are regions of birth and  emigration (near saddles)
and regions of immigration and death (within the states).
There is only a small global death rate of the order of
the inverse reaction time along the specific path.

Denoting $F({\boldsymbol{x}},{\boldsymbol{v}},t)$ the  density
of walkers, the current distribution is calculated from the average
$
R_i({\boldsymbol{x}},t) = \int {\boldsymbol{dv}}\; {\boldsymbol{v}} \;
F({\boldsymbol{x}},
{\boldsymbol{v}},t)
$.
One can  check that 
the $R_i$ then satisfy (\ref{onef}), as desired.
 This algorithm can be seen as
 a  Diffusion Monte Carlo as applied to (\ref{onef}) (see    Refs.
 \cite{Fahy91} and \cite{Fantoni00}).

\begin{figure}
\centering \includegraphics[width=70mm]{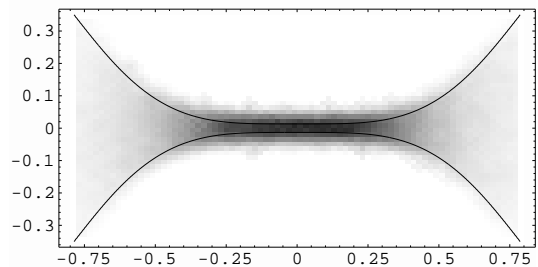}
\caption{Locating an entropic barrier with the vector walkers
  The potential
is flat between the two curves, and has steep harmonic walls.
 Darker regions
correspond to larger values of the {\em barrier resistance} (see text).}
\label{entr}
\end{figure}

Once an unnormalized transition current ${\boldsymbol J}$ is known,
the reaction time is obtained for example through:
\begin{equation}
\tau_{activ} = \frac{\int d {\boldsymbol x}\; 
e^{ \beta E} | {\boldsymbol J}|^2}{ \int d {\boldsymbol x}\; 
e^{\beta E}  ({\mbox{div}} {\boldsymbol J})^2}
\label{barrres}
\end{equation}
where $\beta=\frac{1}{T}$. The {\em barrier resistance}  $e^{ \beta E} | {\boldsymbol J}|^2$
is large in (and serves to indicate) barriers, while 
$e^{ \beta E}  ({\mbox{div}} {\boldsymbol J})^2$ is concentrated and 
essentially constant within the states involved in the passage~\cite{Gaveau98}.
In Fig. \ref{entr} we show the barrier resistance, as obtained from
the present algorithm, for a purely entropic barrier.

Let us conclude by mentioning that sign problems are not
{\em a priori} excluded here. They consist in the appearance of neighboring walkers with opposite vectors.
For very low temperatures, and in particular for the 
evaluation of saddle-points, one can show \cite{Tanaseu} that this poses no problem. 
Even at finite temperatures walker cancellations turn out not to be serious. However,
one can always  decimate the 
walkers, lumping them together according to the mean local orientation.

{\bf Transition path sampling with saddle-seeking paths.}
One can also use these ideas  to improve transition path sampling 
\cite{Dellago98,Bolhuis02,E02}.
This family of techniques involves sampling 
over trajectories, weighted with the appropriate
 action. In order to obtain information on reaction paths, it is necessary to
force the ends of the trajectory to be
 at either side of a barrier --- otherwise the 
trajectory collapses to a globular shape around  one minimum.

If we wish the trajectory to  seek for saddles, we need to add the effect of
 the second term on the rhs of 
(\ref{onef}). A simple calculation
 shows that this involves adding to the usual action 
a Lyapunov term equal to the logarithm of the largest 
eigenvalue of $U(t)$  defined 
from~\footnote{Note that this differs from the usual Lyapunov exponent that is defined 
considering  $U^\dag U$ ---
but one can that the eigenvalues of $U$ are on average also positive.}:
\begin{equation}
{\dot U}_{ij}= - E_{ik} U_{kj}
\end{equation}
(Adding in general the  $k$-th eigenvalue 
makes the path seek the saddles with $k$ unstable directions).
 
\begin{figure}
\centering \includegraphics[width=70mm]{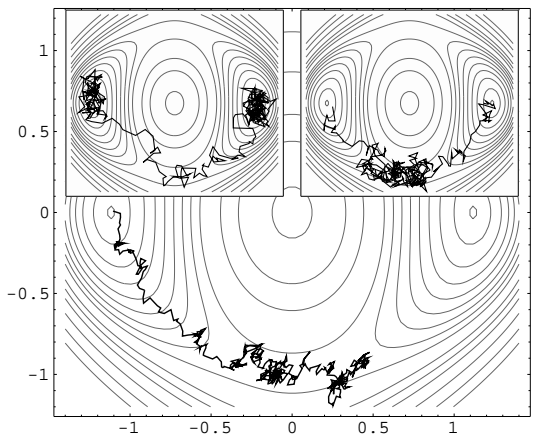}
\caption{Above: two-ended transition path sampling. Left: Langevin weight, right: Langevin
$+$ Lyapunov weight. Below: Langevin $+$ Lyapunov action with
  only one end fixed  while the other end is around a saddle.
The potential is as in Fig. \ref{arrowsnapchan} \cite{Dellago98}.}
\label{chan}
\end{figure}

One can show that the path obtained with the Lyapunov term is concentrated 
with a density proportional to the barrier resistance
$e^{\beta E} | {\boldsymbol J}|^2$ (cfr. Eq. (\ref{barrres})):  
thanks to this modification, if the trajectory is not constrained in 
any way, {\em it will spend its time near a barrier (even if  entropic),
and not in a state} (see Fig. \ref{chan}), although the value of the action remains the same
as for a pure Langevin weight. 
Thus the sampling can take better advantage of the 
regions that are difficult for the passage, also eliminating the problem of
the large fluctuations associated with the jump time. 
 If instead one end is tied to a minimum,
the trajectory leaves the
 minimum and collapses to a globular shape around a nearby saddle
(see Fig. \ref{chan}): this can be used as a
 strategy to explore exit paths from a known 
configuration.
The  transition times in this setting can be calculated adapting the methods in
\cite{Dellago98}.

{\bf  \small THEORETICAL BACKGROUND}

To show these results, and to place them in a wider context, we introduce the supersymmetric formalism.
Let us start by  artificially `completing the square'  in (\ref{HFP}) as follows:
we introduce the $N$ fermion
creation and annihilation operators $a_i$ and $a^\dag_i$, with
anticommutation relations $[a_i,a^\dag_j]_+=\delta_{ij}$, and 
define charges as:
\begin{equation}
 {\bar Q}=-i a^\dag_i (T \partial_i +E_{i}) \;\;\; ; \;\;\; 
Q=-i T a_i \partial_i 
\end{equation}
Both  operators ${\bar Q}$ and $Q$ commute with $H$, are nilpotent 
${\bar Q}^2=Q^2=0$, and:
\begin{equation}
H = \frac{1}{T} ({\bar Q}+Q)^2= H_{\mbox{{\tiny FP}}} +  E_{ij} a^\dag_j a_i 
\label{HSUSY}
\end{equation}
We consider a wave function $|{\boldsymbol \psi}\rangle$ in the enlarged space and:
\begin{equation}
\frac{d|\boldsymbol{\psi}\rangle}{dt}=
 - H|{\boldsymbol{\psi}}\rangle
\label{Seq}
\end{equation}
which yields a separate evolution for each fermion number subspace.
The zero-fermion subspace is just the original space of probability 
distributions and (\ref{Seq}) is the Fokker-Planck equation 
 (\ref{HFP}) for them.
The one-fermion subspace is made of $N$-component 
functions $R_i(\boldsymbol{x}) a^\dag_i |\rangle $ ($|\rangle $ the fermion vacuum)
 and (\ref{Seq}) reduces to (\ref{onef}) for them.
 More generally, the  $k$-fermion
subspace is spanned by  $N!/k!(N-k)!$ functions.

As is well known, $H_{\mbox{{\tiny FP}}}$ can be taken to a Hermitian form 
$ H_{\mbox{{\tiny FP}}}^h = e^{\beta E/2}  H_{\mbox{{\tiny FP}}}  e^{-\beta E/2}$,
and this is also true of $H$:
\begin{eqnarray}
 H_{\mbox{{\tiny FP}}}^h &=&  
\frac{1}{T}\sum_i 
\left[-T^2 \partial^2_i + 
\frac{1}{4} E_{i}^2 - \frac{T}{2} E_{ii}
\right] \label{hermi} \\
H^h  &=& e^{\beta E/2}  H  e^{-\beta E/2} =  H_{\mbox{{\tiny FP}}}^h +   E_{ij} a^\dag_j a_i
\end{eqnarray}
The eigenvalue equations are:
\begin{equation}
H \boldsymbol{ | \psi^R \rangle} = \lambda 
\boldsymbol{ | \psi^R \rangle}   \;\; ; \;\;
 H^h \boldsymbol{ | \psi^h \rangle}  
= \lambda   \boldsymbol{|\psi^h \rangle   } \nonumber   
\end{equation}
where
\begin{equation}
\boldsymbol{ | \psi^R \rangle} =  e^{-\beta E/2} 
\boldsymbol{ | \psi^h \rangle}    
\end{equation}

\begin{figure}
\centering \includegraphics[width=7cm,totalheight=4cm]{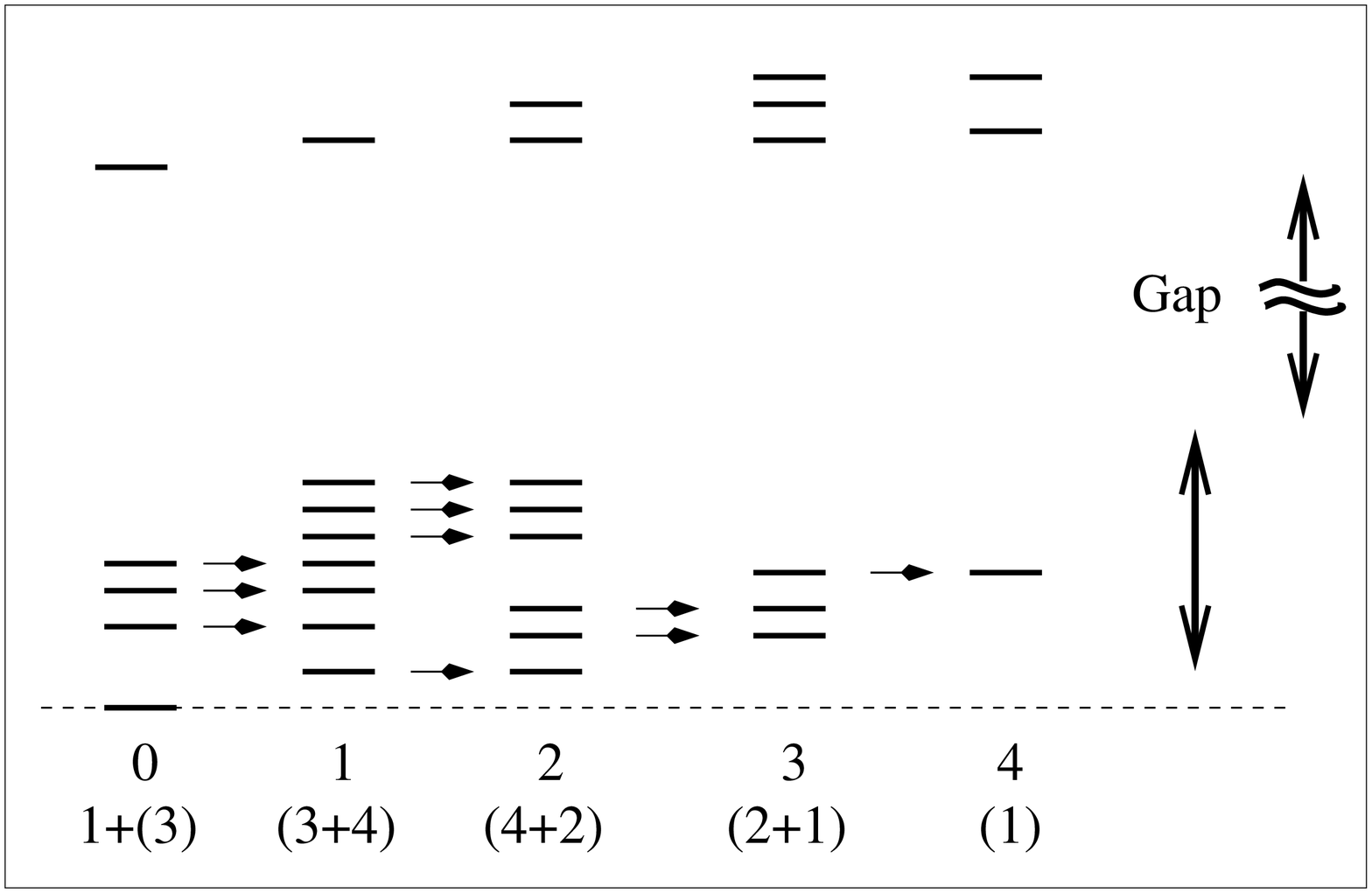}
\caption{
Morse theory. The arrows indicate the action 
of ${\bar{Q}}$. The gap in the spectrum reflects timescale separation.
The numbers between brackets indicate the number of states below the
gap  for each fermion number. The Morse inequalities can be read from
the picture. 
}
\label{Spectra}
\end{figure}

Now, it is easy to see that applying $Q$ to any eigenvector  
$\boldsymbol{ | \psi^R \rangle}$ we get either a degenerate
eigenvector with one  less fermion, 
 or zero. Similarly,  applying ${\bar Q}$
 we get either a degenerate eigenvector with one more fermion
 or zero.
Clearly, both $Q$ and  ${\bar Q}$ annihilate the Gibbs measure. 
The spectrum is thus organized as in Fig \ref{Spectra}.

The subspaces with non-zero fermion numbers
are at this stage artificial. {\em The main result we need  here
is that small-eigenvalue (right)  eigenvectors with one fermion encode the transition currents 
between states. On the other hand, small-eigenvalue $k$-fermion eigenvectors in the
Hermitian base are peaked in all the saddle points with exactly $k$ unstable directions.
}

{\bf States.}
Before discussing transition distributions, we need to specify what we
understand in general by `states'.
In cases in which there is a timescale separation $\tau_{fast} \ll \tau_{activ}$,
whatever its origin, the Fokker-Planck
spectrum has a gap $\sim \tau_{fast}^{-1} - \tau_{activ}^{-1}$.
Suppose there are $K$ eigenvalues `below the gap'.
 An important and intuitive
result due to   Gaveau and Schulman
\cite{Gaveau98}
is that using 
linear combinations of them we can construct $K$ distributions that are 
 essentially Gibbsean in $K$ disjoint regions, and elsewhere negligible. 
 In other words,  metastable states (up to a lifetime
 $\tau_{activ}$) are linear combinations of
states below the gap, and conversely.
All these statements become sharper the larger the gap.
The case of small temperatures is emblematic: the `states' are simply
 narrow  distributions (typically
of width $\sqrt T$) peaked
around local minima. This can be shown immediately from the low $T$ 
expansion of (\ref{hermi}).

{\bf Reaction Paths.} Let us now sketch a proof of the fact that
 {\em one-fermion eigenvectors below the gap give the 
reaction-path distribution between states, and their times}.
This 
means 
that Eq. (\ref{onef}) on times larger than $\tau_{fast}$ gives the 
reaction paths, {\em and this in times
much smaller than the passage-times $\tau_{activ}$ themselves}.

Consider for simplicity the low-temperature
situation, in which there are $K$ minima separated by energy barriers, and
 a Langevin evolution starting from each minimum
with a certain probability.
If the time is of the order of the inverse of the largest passage time between any
of the
 $K$ states, there will be a current
 leaking through this saddle, and negligible flowing 
through all other paths (which take exponentially longer times). 
Next, consider times enough to allow the second fastest passage,
there will be current flowing only through it:   the faster reaction
has already taken place (the states involved have mutually
thermalized)
and the slowest reactions-currents are exponentially smaller.
If we continue this reasoning $K$ times, each time allowing for a new 
activation, we obtain a sequence of  $K$ different
linear combinations 
$\phi_K(\boldsymbol{x}),...,\phi_1(\boldsymbol{x})$
 of the states below the gap,
 the last one $\phi_1(\boldsymbol{x})$ the Gibbs measure.

 Consider next
the $K-1$ 
 one-fermion right eigenvectors   obtained acting with  ${\bar Q}$ on  
the  
$\phi_2(\boldsymbol{x}),...,\phi_K(\boldsymbol{x})$:
\begin{equation}
 {\bar Q} \phi_a^L  = (T \partial_i+E_i)a^\dag_i \phi_a^L=J_i^a(x)  a^\dag_i |\rangle
\label{form}
\end{equation}
From the preceding argument,
they represent exactly $K-1$ independent passage current 
distributions. Hence, in general each individual one-fermion eigenvector
 below the gap obtained as in (\ref{form})
represents a passage current distribution
with sources and sinks in the states (themselves defined as above).
In the low temperature case, the distribution is non-negligible along a
tube following  ascending and descending  gradient lines,
 passing through a barrier.
 The direction of
$J_i^a(x)$ is parallel to the tube.
The converse is also true:
 each passage between any two states is
a linear superposition of these $K-1$ currents.

In Fig. \ref{Spectra}, we see that there are also 
one-fermion eigenvectors that are not the result of acting
 with ${\bar Q}$ as in (\ref{form}). They are characterized by being annihilated by
$Q$, and this implies 
that they correspond to divergenceless current distributions.
 They are `tours', for example
an activated process leading from a minimum
 to itself through a saddle (think for example
of a tilted Mexican hat). They are automatically distinguished in this
formalism: for example (\ref{barrres}) yields infinite timescales for them.

{\bf Hermitian base: saddles and Morse Theory.}
One way to convince oneself  that the formalism is a natural one
is to see how it yields quite simply
the topological relations between saddles (Morse theory).
Let us review briefly the derivation in \cite{Witten82},
specializing
to the case of the topology of ordinary flat space and bounded systems.
Consider the low temperature   
(harmonic) spectrum  of (\ref{hermi}): it is easy to show that
 $k$-fermion eigenvectors that have zero
eigenvalue to leading order, are narrow Gaussians sitting
 on saddles with $k$ unstable directions.
This one-to-one relation
between eigenfunctions `below the gap'  
{\em in the Hermitian base} and saddles
 implies, as can be seen from Fig. \ref{Spectra}  that there are relations between
 the numbers of saddles with successive numbers of unstable directions:
 $(1+n_1)$, $(n_1+n_2)$, ... , $(n_{N-1}+n_N)$.  The fact that $n_a>0$ 
constitute the strong Morse inequalities for our case.

{ \bf \small CONCLUSIONS}

We described a family of methods to find transition paths and saddle points
that, to the best of our knowledge, are
quite different from the standards ones. The statement
of the algorithms is simple, but the inspiration for them -- and indeed the proof that
they work -- is based on the supersymmetric quantum mechanics construction, 
and its connection to Morse theory. Although we could not provide here 
 many details of the derivations, and we have not yet performed   
tests in really hard cases,  we hope we have argued convincingly that 
the tools are quite relevant for the problem~\cite{Tanaseu}.

{\bf Acknowledgments.} We wish to thank N. Mousseau, W. Krauth, F. Schm\"user and D. Wales for useful discussions.

\end{document}